\documentclass{PoS}

\title{Jet/medium interactions at large scales}

\ShortTitle{Jet/medium interactions}

\author{\speaker{Pol Bordas}\thanks{P.B has been supported by grant DLR 50 OG 0601 during this work. P.B acknowledges support by the Spanish DGI of MEC under grant AYA2007-6803407171-C03-01. P.B also acknowledges the excellent work conditions at the \textit{INTEGRAL} Science Data Center.}\\
        Institut f\"ur Astronomie und Astrophysik, Universit\"at T\"ubingen, Sand 1, 72076 T\"ubingen, Germany\\
        {\textit{INTEGRAL}} Science Data Centre, Universit\'e de Gen\`eve, Chemin d'Ecogia 16, CH--1290 Versoix, Switzerland\\
        E-mail: \email{pol.bordas@uni-tuebingen.de}}

\author{V. Bosch-Ramon\thanks{V.B-R. acknowledges support of the Spanish MICINN under grant
AYA2007-68034-C03-1 and FE\-DER funds.
V.B-R. also acknowledges the support of the European Community under a Marie Curie Intra-European 
fellowship.}\\
        Dublin Institute for Advanced Studies, 31 Fitzwilliam Place, Dublin 2, Ireland.\\
        E-mail: \email{valenti@cp.dias.ie}}
\author{M. Perucho\thanks{M.P acknowledges support from the Spanish MEC and the European Fund for Regional Development 
through grants AYA2010-21322-C03-01, AYA2010-21097-C03-01 and CONSOLIDER2007-00050, and
from the ``Generalitat Valenciana'' grant 
``PROMETEO-2009-103''. M.P acknowledges support from MICINN through a
``Juan de la Cierva'' contract.}\\
Departament d'Astronomia i Astrof\'isica. Universitat de Val\`encia. C/ Dr. Moliner 50, 46100 Burjassot (Val\`encia), Spain\\
        E-mail: \email{Manel.Perucho@uv.es}}

\abstract{High energy emission could be produced in the interaction sites of both galactic and extragalactic jets with the surrounding medium.  We have developed an interaction model that accounts for the continuous injection of relativistic electrons in the forward, reverse and recollimation shocks. We also performed hydrodynamical simulations to establish the physical properties in both type of systems. The resulting non-thermal emission is predicted assuming  different values for the jet power, the external mass density and the source age for both FR-I galaxies and galactic microquasars. The obtained fluxes are compared to current instrument sensitivities at radio, X-ray and gamma-ray bands. We study in detail the connection between hard X-ray synchrotron radiation and gamma-ray emission and use the INTEGRAL, Fermi and AGILE, and Cherenkov telescopes capabilities to test it.}

\FullConference{8th INTEGRAL Workshop: The Restless Gamma-ray Universe Integral2010,\\
		September 27-30, 2010\\
		Dublin Ireland}

\begin{document}

\section{Introduction}

The jets of Fanaroff-Riley galaxies (of type FR-I and FR-II, \cite{fr74}) deliver kinetic energy to the surrounding interstellar and intergalactic medium (ISM and IGM respectively) at a rate between $\sim 10^{42}$ and $\sim 10^{46}\,\rm{erg~s}^{-1}$. The ejections of microquasars ($\mu$Q), on the other hand, can also inject large amounts of energy into the ISM, at a level $\sim 10^{37}-10^{39}$~erg~s$^{-1}$. Both of $\mu$Q and FR-I jet/medium interactions could be strong enough to accelerate particles and produce non-thermal radiation \cite{br09, br10}. Extended X-ray emission from the jets and/or lobes of the radio galaxies 3C~15 \cite{ka03}, Cen~A \cite{cr09}, Fornax~A \cite{fe95} and M87 \cite{ka05} have been observed with \textit{Chandra}. Furthermore, the {\it Fermi} Collaboration recently reported on the extended GeV emission from Cen~A \cite{ab10a}.  Efficient particle acceleration is therefore taking place at least in some FR-Is. In the case of $\mu$Qs, evidence of particle acceleration is found in the non-thermal radio and/or X-ray emission observed e.g. in SS~433 \cite{zealey80},  XTE~J1550$-$564 \cite{corbel02} and Cir~X-1 \cite{tudose06}), and perhaps also in GRS~1915+105 \cite{Kaiser04} and LS~I~+61~303 \cite{paredes07}. In this work we explore the leptonic non-thermal emission expected in both $\mu$Q and FR-I interaction scenarios, using the results from numerical simulations coupled to a radiation model for the jet shocked material (reconfinement region and cocoon) and the ambient material shocked by the bow shock (the shell). Our results are used to make predictions for the broadband non-thermal fluxes (from radio to $\gamma$-rays) and we compare them  with current and future instrument capabilities.

\section{Jet/medium interaction model}

Two symmetric jets emerge from the central source, expanding freely until they are recollimated when their ram pressure equals that of the surrounding cocoon. Further on, the jet is decelerated once the accumulated mass of the swept up ISM/ICM gas becomes similar to that carried by the jet. A forward shock is produced propagating into the external medium, whilst the shocked matter from the jet inflates the cocoon. In the FR-I scenario, the jet gets disrupted before reaching the termination regions, and we take the jet recollimation shock as the cocoon particle accelerator. In the $\mu$Q scenario the jet is assumed to remain undisrupted, and a strong reverse shock is formed near the jet head. In our model the mass density and pressure of the shocked regions are taken homogeneous, although they can evolve with time. We assume the presence of a randomly oriented magnetic field $B$ in the downstream regions, derived taking the magnetic energy density to be $\sim$~10~\% of the internal energy density. In each shocked region, the fraction of kinetic power  transferred to non-thermal particles is taken to be $\sim$~1~\%. We consider the CMB and the radiation field energy density from the central engine (the companion  star in the case of $\mu$Q and the galaxy nucleous in the case of FR-Is), although the latter is only important at the very initial stages of evolution. We adopt a power-law spectral distribution ($N(E)=K\,E^{-2}$, with $p = 2$ and $K$ such that $\int E\,N(E)\,dE = 0.1 \times Q_{\rm jet}$) for the leptons injected at the reconfinement, bow and reverse shock fronts (we remark that the latter is only considered in the undisrupted jets of $\mu$Qs). Maximum energies are calculated equating the energy gain to synchrotron, relativistic Bremsstrahlung, Inverse Compton (IC) and adiabatic losses. Differently evolved populations injected all along the source age are considered at a given $t_{\rm src}$. For the synchrotron losses ($t_{\rm syn}\approx 4\times 10^{12}\,(B/{\rm 10 \mu~G})^{-2}\,(E/{\rm 1 TeV})^{-1}\,{\rm s}$), we use the magnetic field considered above for each interaction region. Relativistic Bremsstrahlung is calculated accounting for the densities $n$ in the downstream regions ($t_{\rm rel.br}\sim 10^{18}\,(n/10^{-3}\,{\rm cm}^{-3})\,{\rm s}$). To compute IC losses ($t_{\rm IC}\approx 1.6\times 10^{13}\,(u_{\rm rad}/10^{-12}{\rm erg~cm}^{-3})^{-1}\,(E/{\rm 1 TeV})^{-1}\,{\rm s}$), we consider the total radiation field energy density $u_{\rm rad}$. Adiabatic losses, $\dot{E} \approx [v/r]\,E$, are computed from the size $r$ and the expansion velocity $v$ of the emitters. Escape losses are also consideredd by taking the gyroradius of the most energetic particles equal to the accelerator size \cite{Hillas1984}.

\section{Hydrodynamical simulations}

%
%

Hydrodynamical simulations have been performed to further study the interaction of both $\mu$Q and FR-I jets with their surroundings. A two-dimensional finite-difference code based on a high-resolution shock-capturing scheme has been used, which solves the equations of relativistic hydrodynamics written in conservation form. The reader is referred to \cite{pe+05,pm07} for further details on the simulation code (see also \cite{br09}). For the $\mu$Q scenario, the jet is injected at a distance of $10^{18}$~cm from the compact object, with initial radius $ = 10^{17}$~cm. Both the jet and the ambient medium are characterized with an adiabatic exponent $\Gamma=5/3$. The number density in the ambient medium is $n_{\rm ISM}=0.3$~cm$^{-3}$. The velocity of the jet at injection is $0.6\,c$, its number density $n_{\rm j}=1.4\times 10^{-5}\,\rm{cm^{-3}}$, and its temperature $T\sim 10^{11}$~K. These parameter values result in a jet power $Q_{\rm jet}=3\times 10^{36}$~erg~s$^{-1}$. For the numerical simulations of FR-I sources, the jet is injected in the numerical grid at $500\,\rm{pc}$ from the active nucleus, with a radius of $60\,\rm{pc}$. The ambient  medium is composed by neutral hydrogen. Its profile in pressure, density and temperature  includes the contribution from a core region and from the galaxy group, which dominates at large distances. The initial jet velocity is $\sim 0.87\,c$, its temperature $\sim 4\times10^9\,\rm{K}$, and a density and pressure ratio with respect to the ambient are $\sim 10^{-5}$ and $\simeq 8$, respectively, resulting in a jet kinetic luminosity $Q_{\rm j}=10^{44}$~erg~s$^{-1}$.  
%


\section{Non-thermal emission from $\mu$Q}

 Figure~1 show the spectral energy distribution (SED) for the shell, the cocoon and the jet reconfinement regions for the different set of parameters explored. The emission of only one jet is accounted for. Synchrotron emission is the channel through which the highest radiation output is obtained, with bolometric luminosities up to $\sim~10^{33}$~erg~s$^{-1}$ for powerful ($Q_{\rm jet}= 10^{37}$~erg~s$^{-1}$) systems. At high and very high energies (HE and VHE, respectively), IC emission is the dominant process in the cocoon and reconfinement regions, reaching a few $\times 10^{30}$~erg~s$^{-1}$, while in the shell relativistic Bremsstrahlung dominates at this energy range, with luminosities up to $\sim~10^{32}$~erg~s$^{-1}$. Notable differences are found in the reported SEDs when varying the source age $t_{\rm MQ}$ from $10^{4}$~yr to $10^{5}$~yr. For older sources, the shell and cocoon are located at larger distances from the companion star, $u_{\rm rad}$ decreases and the IC contribution gets slightly lower. Relativistic Bremsstrahlung emission is also larger for older sources. Higher values of $n_{\rm ISM}$ make the jet to be braked at shorter distances from the central engine. The interaction regions have higher $u_{\rm rad}$ from the companion star, and the IC emission is again slightly enhanced. The relativistic Bremsstrahlung emission in the shell zone is also higher for denser mediums, since the luminosity is proportional to the target ion field density, $n_{\rm t} \sim 4\,n_{\rm ISM}$.
Finally, we note that in our model all the non-thermal luminosities scale roughly linearly with the jet power $Q_{\rm jet}$.

\begin{figure}
\centering 
\includegraphics[width=1.0\textwidth]{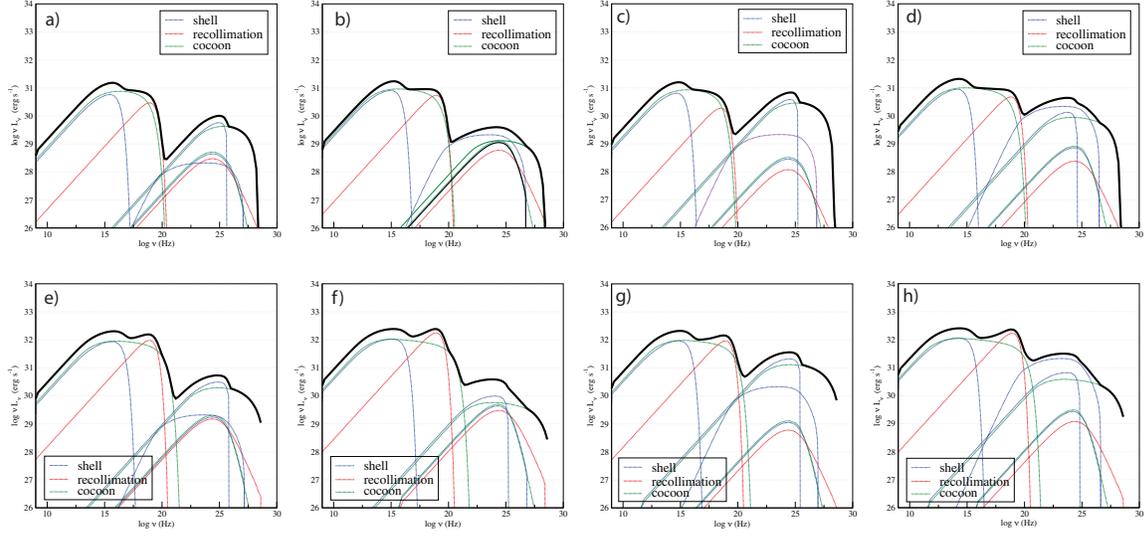}

\begin{footnotesize}\caption{Spectral Energy Distribution of the non-thermal emission produced in the $\mu$Q jet/ISM interaction. The contribution from the shell, the cocoon and the recollimation regions are shown in blue, red and green dashed lines, respectivey, whilts the overall emission is represented by the thick black continuous line. For the shell, synchrotron, IC and reativistic Bresstrahlung have been considered. For the recollimation and cocoon regions, synchrotron and IC are accounted for. Upper and lower panels correspond to a ISM particle density $n_{\rm ISM} = 0.1$~cm$^{-3}$ and 1~cm$^{-3}$, respectively.  Panels a), b), e), and f) show the a source age of $t_{\rm src} = 10^{4}$~yr, whilst c), d), g) and h) correspond to $t_{\rm src} = 10^{5}$~yr. Cases a), c), e) and g) correspond to a jet power $Q_{\rm jet} = 10^{36}$~erg~s$^{-1}$,  whilst b), d), f) and h) are for $Q_{\rm jet} = 10^{37}$~erg~s$^{-1}$. See text for details on the rest of parameters used in the radiative model.} \end{footnotesize}
\label{SEDS_mq_TMQ4_TMQ5}
\end{figure}

\section{Non-thermal emsission from FR-I sources}

\begin{figure}[t]
\centering 
\includegraphics[width=1.0\textwidth]{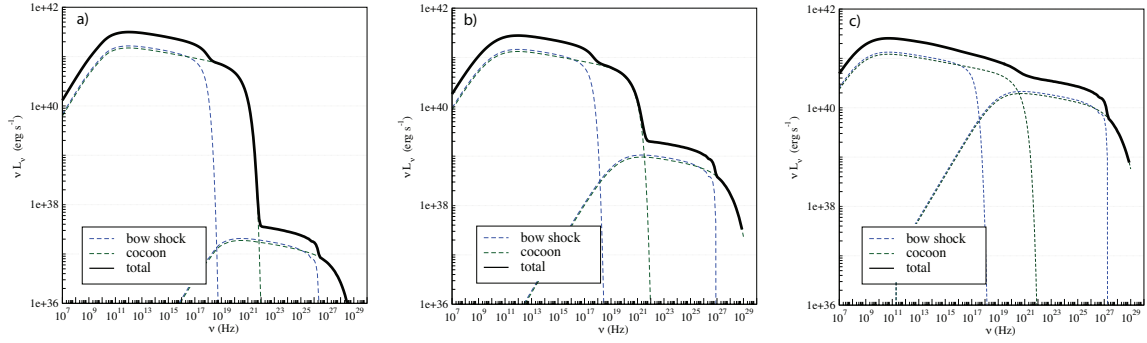}
\caption{SEDs of the non-thermal emission produced in the interaction of FR-I jets with their surroundings. Panels a), b) and c) correspond respectively to a source age $t_{\rm src} = 10^{5}, 3 \times 10^{6}$ and $10^{8}$~yr. Blue dashed lines show the contribution from the shell region and green dashed lines that of the cocoon. The overall emission is marked with a thick solid black line.} 
\label{SED_FRI_totes}
\end{figure}

The SEDs for the cocoon and the shell regions at $t_{\rm src}=10^{5}$, $3 \times 10^{6}$
and $10^{8}$~yr, are shown in Fig.~\ref{SED_FRI_totes}. The obtained radio and X-ray synchrotron luminosities in
both regions are at the level of $2\times 10^{41}$~erg~s$^{-1}$.
The synchrotron break frequency, corresponding to the electron energy at which $t_{\rm syn}(E)\approx t_{\rm src}$, and the
highest synchrotron frequency, $\nu_{\rm syn~max}\propto B\,E_{\rm max}^2$, are shifted down for older sources. The former 
effect makes the radio luminosity to increase at the late stages of the evolution of both the cocoon and the shell, whereas
the latter decreases the X-ray luminosity in the shell due to the decrease of $\nu_{\rm syn~max}$ with time. The slightly
different conditions in the shell yield a higher break frequency, which implies a factor  $\sim 2$ lower radio luminosity in
this region compared to that of the cocoon. The IC luminosity grows as long as this process becomes more efficient compared to synchrotron and adiabatic cooling. The cocoon and the shell have similar HE luminosities, but the cocoon is few times brighter at VHE than the shell. In both regions the bolometric IC luminosities grow similarly with time, reaching $\sim 10^{42}$ and $10^{41}$~erg~s$^{-1}$ at HE and VHE, respectively.

\section{Discussion}

The obtained radio fluxes in a $\mu$Q scenario would imply a flux density of $\sim 150$~mJy at 5~GHz for a source located at $\sim$~3 kpc. The
emitting size would be of a few arcminutes, since the electron cooling timescale is longer than the source lifetime and they can fill the whole
cocoon/shell structures. Considering this angular extension and taking a radio telescope beam size of $10''$, radio emission at a level of $\sim
1$~mJy~beam$^{-1}$ could be expected. In X-rays, we find a bolometric flux in the range 1--10~keV of $F_{\rm 1-10 keV}$ $\sim 2 \times
10^{-13}$~erg~s$^{-1}$~cm$^{-2}$. The electrons emitting at X-rays by synchrotron have very short time-scales, and the emitter size cannot be
significantly larger than the accelerator itself. Although the X-rays produced in the shell through relativistic Bremsstrahlung are expected to be
quite diluted, the X-rays from the cocoon would come from a relatively small region close to the reverse shock, and could be detectable by {\it
XMM-Newton} and {\it Chandra} at scales of few arcseconds. At hard X-rays, \textit{INTEGRAL} could detect the high energy tail of the particle population, although its moderate angular resolution would make difficult to resolve the sources. At HE, the flux between 100 MeV and 100 GeV is $F_{\rm
100~MeV<E<100~GeV}\sim 10^{-14}$~erg~s$^{-1}$~cm$^{-2}$, below the \textit{Fermi} sensitivity, while the integrated flux above 100~GeV is $F_{\rm E>100GeV}\sim 10^{-15}$~erg~s$^{-1}$~cm$^{-2}$, also too low to be detectable by current Cherenkov telescopes. Taking into account the rough linearity between $Q_{\rm jet}$, $n_{\rm ISM}$, $t_{\rm src}$ and $d^{-2}$ with the gamma-ray fluxes obtained, sources with higher values of these
quantities than the ones used here may render the $\mu$Q jet termination regions detectable by current and next future gamma-ray facilities.


For FR-I sources radio fluxes as high as $\sim 10^{-12}\,(d/100~$Mpc$)^{-2}$~erg~cm$^{-2}$~s$^{-1}$ or $\sim 10$~Jy at 5~GHz
from a region of few times $10'\,(d/100~$Mpc$)^{-1}$ angular size are obtained. The properties of the radio emission from the interaction
jet-medium structure are comparable with those observed for instance in 3c~15 \cite{ka03}, in which fluxes of a few $\times
10^{-14}$~erg~cm$^{-2}$~s$^{-1}$ ($d\sim$~300~Mpc) are found. At X-rays, the dominant emission comes from the cocoon, with fluxes $\sim 10^{-13}\,(d/100~$Mpc$)^{-2}$~erg~cm$^{-2}$~s$^{-1}$, although limb brightening effects may increase the shell detectability. The lifetime of X-ray synchrotron electrons, $\sim 10^{11}\,{\rm s}$, is much shorter than in radio, and $\ll t_{\rm src}$ as well, so their radiation may come mostly from the inner regions of the cocoon or the bow-shock apex. The total non-thermal X-ray flux is roughly constant for the explored range of $t_{\rm src}$, although the shell contribution decreases significantly with time. At HE and VHE energies, the obtained fluxes are similar for both the cocoon and the shell, although the latter shows a lower
maximum photon energy. Gamma-ray emission increases
with time mainly due to the increasing efficiency of the CMB IC channel.
The obtained fluxes, for a source with $t_{\rm src}\sim 10^8$~yr, are around $\sim 10^{-12}\,(d/100\,{\rm
Mpc})$~erg~cm$^{-2}$~s$^{-1}$. At HE, such a source may require very long exposures to be detected. Nevertheless, {\it Fermi}
has recently detected some FR-I galaxies at a few
hundred Mpc distances, including the extended radio lobes of Cen~A (at a distance $\sim 4$~Mpc), presenting fluxes similar to those predicted here. At VHE, the fluxes would be detectable by the current
instruments, although the extension of the source, of tens of arcminutes at 100~Mpc, and the steepness of the spectrum above
$\sim 100$~GeV, may render its detection difficult. In the case of Cen~A, detected by HESS \cite{ah09}, the emission seems to
come only from the core, but this is expected given the large angular size of the lobes of this source, which would dilute
its surface brightness, requiring very long observation times to render the source detectable. Finally, we note that the fluxes showed above strongly depend on the non-thermal luminosity fraction going to non-thermal particles, for which we use a quite
conservative value $=0.01$. In the case of a source able to accelerate particles at a higher efficiency at the interaction shock fronts, then
the expected non-thermal fluxes would be enhanced by a similar factor.

\end{document}